\documentclass[aip,cha,amsmath,amssymb,reprint,numerical]{revtex4-1}

\usepackage{graphicx}
\usepackage{dcolumn}
\usepackage{bm}
\usepackage{subfigure}
\usepackage{color}
\usepackage{verbatim}

\begin{document}

\title{A perturbative approach to Lagrangian flow networks}

\author{Naoya Fujiwara}
\affiliation{Center for Spatial Information Science, The University of Tokyo, 
5-1-5 Kashiwanoha, Kahshiwa-shi, Chiba, 277-8568, Japan}

\author{Kathrin Kirchen}
\affiliation{Research Domain IV -- Transdisciplinary Concepts and Methods, Potsdam Institute for Climate Impact Research, Telegrafenberg A31, 14473 Potsdam, Germany}
\affiliation{Mathematical Institute, University of Bonn, Endenicher Stra{\ss}e 60, 53115 Bonn, Germany}
\affiliation{Department of Geology and Environmental Science, The Kenneth P. Dietrich School of Arts and Sciences, University of Pittsburgh, 4107 O'Hara Street, Pittsburgh, PA 15260-3332, United States}

\author{Jonathan F. Donges}
\affiliation{Research Domain I -- Earth System Analysis, Potsdam Institute for Climate Impact Research, Telegrafenberg A31, 14473 Potsdam, Germany}
\affiliation{Stockholm Resilience Centre, Stockholm University, Kr\"aftriket 2B, 11419 Stockholm, Sweden}

\author{Reik V. Donner}
\affiliation{Research Domain IV -- Transdisciplinary Concepts and Methods, Potsdam Institute for Climate Impact Research, Telegrafenberg A31, 14473 Potsdam, Germany}

\date{\today}

\begin{abstract}
Complex network approaches have been successfully applied for studying transport processes in complex systems ranging from road, railway or airline infrastructure over industrial manufacturing to fluid dynamics. Here, we utilize a generic framework for describing the dynamics of geophysical flows such as ocean currents or atmospheric wind fields in terms of Lagrangian flow networks. In this approach, information on the passive advection of particles is transformed into a Markov chain based on transition probabilities of particles between the volume elements of a given partition of space for a fixed time step. We employ perturbation-theoretic methods to investigate the effects of modifications of transport processes in the underlying flow for three different problem classes: efficient absorption (corresponding to particle trapping or leaking), constant input of particles (with additional source terms modeling, e.g., localized contamination), and shifts of the steady state under probability mass conservation (as arising if the background flow is perturbed itself). Our results demonstrate that in all three cases, changes to the steady state solution can be analytically expressed in terms of the eigensystem of the unperturbed flow and the perturbation itself. These results are potentially relevant for developing more efficient strategies for coping with contaminations of fluid or gaseous media such as ocean and atmosphere by oil spills, radioactive substances, non-reactive chemicals or volcanic aerosols.
\end{abstract}


\maketitle

\begin{quotation}

Perturbation-theoretic methods have found wide applications in many areas of physics, ranging from classical celestial mechanics to quantum physics. At various occasions, they have proven useful for addressing scientific problems where an explicit analytical treatment is not possible, but the system under study can be considered as a minor modification of another problem where such an analytical solution can be obtained. This work combines basic concepts of perturbation theory with Lagrangian flow networks, a novel tool that allows characterizing structural properties of flows by means of complex network theory. Specifically, by means of an eigenvector decomposition of the associated transition matrix (i.e., the spatio-temporally discretized Perron-Probenius operator), it is studied how minor modifications affect the steady-state distribution of passively advected particles in a given flow pattern. One main potential field of application of the developed framework is atmospheric and ocean physics, where the proposed approach may help developing more efficient strategies for coping with contaminations of fluid or gaseous media by oil spills, radioactive substances, non-reactive chemicals or volcanic aerosols (e.g., to determine the most efficient positions for removing the contaminating substances or anticipate their temporary trapping and natural precipitation).

\end{quotation}

\section{Introduction} 
\label{sec:introduction}

In recent years, complex network theory~\cite{ba99,bbv08,newman2010networks} has been successfully applied to studying complex systems serving the purpose of transportation of material or information, ranging from road, railway or airline infrastructures~\cite{Guimera2005,Gastner2006,Barthelemy2006,Chan2011} over biological trnasport and industrial manufacturing~\cite{Buhl2004,Helbing2009} to fluid dynamics~\cite{Rossi2014,ser2015flow,Mheen2013}. The emerging field of climate network analysis has produced relevant insights into the complex dynamics of the Earth's climate system~\cite{tsonis2004architecture,tsonis2006networks,donges2015complex}. The applications of this modern approach to climate science include unraveling the backbone structure of surface ocean currents from surface air temperature data~\cite{dzmk09}, studying transitions in global climate teleconnection structure during different phases of the El Ni\~no Southern Oscillation (ENSO)~\cite{yamasaki2008climate,radebach2013disentangling,wiedermann2016climate} and using causal network reconstruction techniques~\cite{runge2015identifying} to identify different Arctic drivers of mid-latitude winter circulation~\cite{kretschmer2016using}. 

Most climate network studies have followed an Eularian approach in the sense of computing statistical interrelations between fluid dynamical field variables such as temperature or pressure measured at spatially fixed positions (e.g., at fixed grid cells). In contrast to these Eularian climate networks, many recent approaches to describing flow patterns by means of complex network methods~\cite{ser2015flow,rodriguez2016clustering} are Lagrangian in the sense that the focus is on the properties and transition probabilities of particles or small fluid volumes that are advected with the flow. For example, such Lagrangian flow networks have been applied to study the structure of surface ocean currents in the Mediterranean Sea~\cite{ser2015flow}. Related methods are transition networks encoding transition probabilities between discrete or discretized states in general dynamical systems~\cite{Nicolis2005,McCullough2015} and transfer operator techniques with applications to fluid dynamics~\cite{tantet2015early}.

Climate network approaches have been mostly diagnostic~\cite{donges2015complex} (i.e. focussed on studying the \textit{status quo} statistical properties of a given dataset without making inferences about the future) with only a few heuristic attempts at deriving prognostic results from the networks, e.g., for deriving improved predictions of El Ni\~no dynamics~\cite{ludescher2013improved,ludescher2014very} and monsoon onset and withdrawal dates~\cite{stolbova2016tipping}. In this paper, we combine the Lagrangian flow network idea with the established perturbation theory for linear operators \cite{kato2013perturbation}, which has been among other applications widely applied to many problems in quantum mechanics \cite{s85}, to make prognostic inferences on fluid transport dynamics for three distinct scenarios: (i)~efficient absorption (corresponding to particle trapping or leaking), (ii)~constant input of particles (with additional source terms modeling, e.g., localized contamination), (iii)~and shifts of the steady state under probability mass conservation. The proposed methodology presents a first step towards a more efficient control of transport processes in flow systems based on complex network approaches. Potential applications of interest include estimating the spread of contaminations or debris in fluid media~\cite{Povinec2013,Garcia2015} or the effects of various geoengineering proposals on climate dynamics~\cite{caldeira2013science}.

The paper is structured as follows: After introducing the mathematical formalism (Sect.~\ref{sec:formalism}), we illustrate our perturbation theoretical approach numerically by applying it to an idealized two-dimensional Lagrangian chaos model of Rayleigh-B\'enard convection (Sect.~\ref{sec:examples}). Finally, the scope of the proposed methodology is discussed and conclusions are drawn (Sect.~\ref{sec:conclusion}).

\section{Mathematical Theory}
\label{sec:formalism}

\subsection{General setting}

Assume that we have the situation of some physical space, in which we observe the motion of particles. We divide this space into finitely or countably many sub-regions, which will be identified with the nodes of a network, the Lagrangian flow network associated with the given flow field. Having a discrete number $K$ of such spatial elements, we can approximately describe the transport of particles as a Markov chain. Specifically, let $\vec{p}(t)=\left(p_1(t),\dots,p_K(t)\right)^T$ be the vector of residence probabilities of particles at all nodes at time $t$. Given a transition matrix $\mathbf{A}_{\tau}$ as a discretized version of the Perron-Frobenius operator of the underlying flow~\cite{ser2015flow}, which describes the probabilities of transitions between different sub-regions within a discrete time step $\tau$, the time evolution of the residence probabilities is given by
\begin{equation}
\vec{p}(t+\tau)=\mathbf{A}_{\tau}(t)\vec{p}(t).
\label{MarkovChain}
\end{equation}

Under general conditions, the transition matrix can depend on both, time $t$ and the discrete time step $\tau$. In order to keep the considerations in this work as simple as possible, in the following, we will restrict our attention to stationary flows, where the transition matrix does not depend on time. Moreover, we consider $\tau$ as a global pre-defined parameter of the analysis, and simplify our notation by writing $\mathbf{A}:=\mathbf{A}_{\tau}$ from now on.

We recall that any column-stochastic matrix $\mathbf{A}$ has the properties
\[\sum_{i=1}^K A_{ij}=1 \quad \text{and} \quad 0\le A_{ij}\le1\ \; \forall i,j\in\{1,\dots,K\},\]
corresponding to the conservation of probability. We denote $\lambda_i$ ($i=1,\dots,K$) the eigenvalues of $\mathbf{A}$ and sort them in descending order of their real parts as $\lambda_1 \ge \Re \lambda_2\ge \Re \lambda_3\ge\dots\ge \Re \lambda_K\ge 0$. For the sake of simplicity, we assume that the weighted and directed network structure described by $\mathbf{A}$ corresponds to a strongly connected graph, implying that $\lambda_1=1$ is unique.

In a physically plausible case, the motion of particles occurs with finite velocity. Thus, the smaller $\tau$ the more zero entries $\mathbf{A}$ contains, since particles may at most reach a few neighboring sub-regions during such a short time. 

Let $\vec{u}_i^*$ and $\vec{u}_i$ denote the left and right eigenvectors of $\mathbf{A}$ associated with the eigenvalue $\lambda_i$. By assuming non-degeneracy of the associated spectrum (i.e., pairwise distinct eigenvalues of $\mathbf{A}$, $\lambda_i\neq\lambda_j$ $\forall i\neq j$), we have $\vec{u}_i^*\cdot\vec{u}_j = C_i \delta_{ij}$ with $C_i\neq 0$,
where $\delta_{ij}$ is the Kronecker delta. Note that by rescaling of the $\vec{u}_i$, one can always achieve $C_i=1$ for all $i$. However, in what follows, we prefer to consider the more general case.

The (asymptotic) steady state of the Markov chain describing the flow system is given by $\vec{u}_1$. Since $\mathbf{A}$ is a column-stochastic matrix, $\vec{u}_1^*=(1,1,\dots,1)$ is a left eigenvector associated with the eigenvalue $1$.

In practice, we are often interested in a problem that is somewhat more complex than a simple (closed) Markov chain. Specifically, we may want to add a source/sink term $\vec{s}_{\tau}(t)$ to the right-hand side of Eq.~(\ref{MarkovChain}), which describes the amount of particles that are emitted/absorbed in the individual sub-regions described by the nodes of the flow network during a certain time interval $\tau$. Again, in what follows we will ignore a possible time-dependence and write $\vec{s}$ instead of $\vec{s}_{\tau}$ to simplify our notation. 

In this paper, we investigate the case where the transition matrix $\mathbf{A}$ is perturbed and study the resulting effects on the steady state of $\mathbf{A}$. For this purpose, let $\tilde{\mathbf{A}}$ denote the perturbed matrix. In this case, Eq.~$\eqref{MarkovChain}$ can be rewritten as
\begin{equation}
\vec{n}(t+\tau)=\tilde{\mathbf{A}}\vec{n}(t)+\vec{s},
\label{gestoertessystem}
\end{equation}
\noindent
where $\vec{n}(t)=\left(n_1(t),\dots,n_K(t)\right)^T$ contains the number of particles at the each node, and $\vec{n}(t)/N(t)$ with the total number of particles $N(t)=\sum_{i=1}^K n_i(t)$ provides an empirical estimate of $\vec{p}(t)$ if $N(t)$ is sufficiently large. Note that we have to consider $\vec{n}$ instead of $\vec{p}$, since perturbing $\mathbf{A}$ \textit{and} simultaneously adding a source/sink term can relieve the former probability conservation. Looking at Eq.~$\eqref{gestoertessystem}$, we can see that in this case a steady state exists if and only if $(\mathbf{1}-\tilde{\mathbf{A}})$ is invertible. In this case,
\begin{equation}
\vec{n}_{st}:=\left(\mathbf{1}-\tilde{\mathbf{A}}\right)^{-1}\vec{s}
\label{steadystate}
\end{equation} 
\noindent
is the steady-state solution of the system.

More generally, the main idea of expressing a perturbation is to write 
\begin{equation}
\tilde{\mathbf{A}}=\mathbf{A}+\sigma \mathbf{V},
\label{perturbationmatrix}
\end{equation}
\noindent
where $\mathbf{V}$ is the perturbing factor and $\sigma\ll 1$ describes the magnitude of the perturbation. One can easily see, that $\mathbf{V}$ is the reason why the probability condition of $\mathbf{A}$ might not be conserved. 

In the following, we will focus on three cases: (i) having a non-zero probability for the particles to be absorbed at a certain node, but no additional source or sink term, (ii) a constant source term describing generation of particles, together with an absorbing node, and (iii) a perturbation without violation of probability conservation. Note that the difference between an absorption process and a sink (source) is that the efficiency of the former depends on the particle density, while the latter would correspond to removing (injecting) a fixed number of particles per time unit.

\subsection{Absorption problem without supply}
\label{sec:absorption}

Let us first consider the absorption of particles at a single node only as a common example of a perturbation to a passive advection process in a given flow field. Practically, this situation implies that there is a sub-region of the flow domain at which particles are removed with a certain probability. Specifically, let $g_k$ denote the fraction of particles at node $k$ that are removed during a discrete time step $\tau$ (again, we suppress the additional index $\tau$ for brevity). We define $f_k:=g_k/\sigma$ with $\sigma$ being again a parameter that allows us to conveniently scale the magnitude of the perturbation. Then, we can write
\begin{equation}
\tilde{A}_{ij}=(1-\sigma f_k \delta_{jk})A_{ij}
\label{eq:pert_transition}
\end{equation}
\noindent
which according to Eq.~\eqref{perturbationmatrix} corresponds to
\begin{equation}
V_{ij}=-A_{ij}f_k\delta_{jk}.
\label{eq:pert_abs}
\end{equation}

For the following considerations, let us restrict our attention to the case $\vec{s}=\vec{0}$, which means that no particles are additionally emitted into the flow domain. Due to the considered local absorption, for large $t$, the number of particles advected in the flow decays exponentially with time at a rate given by the largest eigenvalue of $\tilde{\mathbf{A}}$ as 
\begin{equation}
\vec{n}(t) \propto (\tilde{\lambda}_1)^t \tilde{\vec{u}}_1
\label{eq:decay_rate}
\end{equation} 
\noindent
with $\tilde{\lambda}_1<1$. 

We now study the effect of the perturbation on the leading eigenvalue and its associated right eigenvector by employing concepts from perturbation theory.
Recall that we have assumed $\lambda_1$ to be non-degenerate. Let $\vec{u}^{(\alpha)}_1,\lambda^{(\alpha)}_1$ denote correction terms of order $\alpha$ in an expansion of the perturbed eigenvector and eigenvalue with respect to the perturbation strength $\sigma$,
\begin{eqnarray}
\tilde{\vec{u}}_1&=&\sum_{l=0}^{\infty} \sigma^l \vec{u}_1^{(l)},
\label{gestoerterev} \\
\tilde{\lambda}_1&=&\sum_{l=0}^{\infty} \sigma^l \lambda_1^{(l)},
\label{gestoerterew}
\end{eqnarray}
\noindent
where $\lambda_1^{(0)}:=\lambda_1=1$ is the leading eigenvalue of the unperturbed system and $\vec{u}_1^{(0)}:=\vec{u}_1$ the associated right eigenvector. In a similar way, we can also express all other eigentriples $\lambda_i, \vec{u}_i^*, \vec{u}_i$. Without loss of generality, we assume in the following that $\vec{u}^{(1)}_1,\vec{u}^{(2)}_1\perp \vec{u}_1$. Note that in the considered case, the term $\lambda^{(1)}_1$ should be smaller than $0$, since otherwise the number of particles would increase with time. Taking the eigenvalue problem for $\tilde{\mathbf{A}}$, 
\[\tilde{\mathbf{A}}\tilde{\vec{u}}_1=\tilde{\lambda}_1\tilde{\vec{u}}_1,\]
and employing Eqs.~$\eqref{gestoerterev}$ and $\eqref{gestoerterew}$ yields
\begin{align}
(\mathbf{A}+\sigma \mathbf{V})\left(\sum_{l=0}^\infty \sigma^l \vec{u}_1^{(l)}\right)=\left(\sum_{l=0}^\infty \sigma^l\lambda_1^{(l)}\right)\left(\sum_{l=0}^\infty \sigma^l \vec{u}_1^{(l)}\right).
\end{align}

Since the previous expression needs to hold uniformly for all possible values of $\sigma$, we can decompose it into separate expressions for the coefficients corresponding to the different powers of $\sigma$, yielding
\begin{align}
&\sigma^0:& \mathbf{A} \vec{u}_1 =& \, \vec{u}_1, \label{ungestoert}\\
&\sigma^1:& \mathbf{A} \vec{u}^{(1)}_1 + \mathbf{V} \vec{u}_1 =& \, \vec{u}^{(1)}_1 + \lambda^{(1)}_1 \vec{u}_1, \label{erstekorrektur}\\
&\sigma^2:& \mathbf{A} \vec{u}^{(2)}_1 + \mathbf{V} \vec{u}^{(1)}_1 =& \, \vec{u}^{(2)}_1 + \lambda^{(1)}_1 \vec{u}^{(1)}_1 + \lambda^{(2)}_1 \vec{u}_1, \label{zweitekorrektur}\\
& & \vdots \nonumber \\
&\sigma^K:& \mathbf{A} \vec{u}_1^{(K)} + \mathbf{V} \vec{u}_1^{(K-1)} =& \, \sum_{l=0}^K \lambda_1^{(l)} \vec{u}_1^{(K-l)}. \label{Ktekorrektur}
\end{align}
Here, Eq.~\eqref{erstekorrektur} gives the first-order correction and Eq.~\eqref{zweitekorrektur} the second-order correction to the perturbed eigenvalue problem, while Eq.~\eqref{ungestoert} again represents the eigenvalue problem of the unperturbed steady-state solution. 

Let us first consider the first-order correction given in Eq.~$\eqref{erstekorrektur}$. By multiplying the left eigenvector $\vec{u}^*_1$ from the left, we find
\begin{align*}
\vec{u}^*_1 \mathbf{A} \vec{u}^{(1)}_1 + \vec{u}^*_1 \mathbf{V} \vec{u}_1 &= \vec{u}^*_1 \cdot \vec{u}^{(1)}_1 + \vec{u}^*_1 \lambda^{(1)}_1 \vec{u}_1,
\end{align*}
which implies that
\begin{align*}
\vec{u}^*_1 \mathbf{V} \vec{u}_1 &= \lambda^{(1)}_1 \vec{u}^*_1 \cdot \vec{u}_1 = C_1 \lambda^{(1)}_1.
\end{align*}
Defining
\begin{equation}
V^{(0)}_{ij}:=\vec{u}^*_i \mathbf{V} \vec{u}_j
\end{equation}
\noindent
and employing Eq.~$\eqref{eq:pert_abs}$, it follows that
\begin{align*}
V^{(0)}_{11}&=\sum_{i,j=1}^K u^*_{1,i} V_{ij} u_{1,j}
=\sum_{i,j=1}^K V_{ij} u_{1,j} \\ 
&=-\sum_{i,j=1}^K A_{ij} f_k \delta_{jk} u_{1,j}
=-\sum_{i=1}^K A_{ik} f_k u_{1,k} = -f_k u_{1,k}
\end{align*}
\noindent
and, hence,
\begin{equation}
\lambda_1^{(1)}=-\frac{f_k}{C_1} u_{1,k}.
\label{eq:1steig_fk}
\end{equation}
\noindent
This means that the first-order correction to the leading eigenvalue is proportional to the absorption rate $f_k$ at the absorbing node $k$.

As the eigenvectors of $\mathbf{A}$ are a basis of $\mathbb{R}^K$, we can represent the associated first-order correction term $\vec{u}^{(1)}_1$ as a linear combination of the $\vec{u}_k$,
\begin{equation}
\vec{u}^{(1)}_1=\sum_{k=2}^K c^{(1)}_k \vec{u}_k,\label{u'}
\end{equation}
\noindent
where we have made use of the fact that $\vec{u}^{(1)}_1\perp \vec{u}_1$. To determine the coefficients $c^{(1)}_k$, we multiply Eq.~$\eqref{erstekorrektur}$ from the left by the left eigenvectors $\vec{u}_k^*$:
\begin{align*}
\vec{u}^*_k \mathbf{A} \vec{u}^{(1)}_1 + \vec{u}^*_k \mathbf{V} \vec{u}_1 &= \vec{u}^*_k \cdot \vec{u}^{(1)}_1 + \lambda^{(1)}_1 \vec{u}^*_k \cdot \vec{u}_1,
\end{align*}
\noindent 
which by applying Eq.~$\eqref{u'}$ leads to
\begin{align*}
\sum_{l=2}^K c^{(1)}_l \vec{u}^*_k \mathbf{A} \vec{u}_l + V^{(0)}_{k1} &= \sum_{l=2}^K c^{(1)}_l \vec{u}^*_k \cdot \vec{u}_l + \lambda^{(1)}_1 C_1 \delta_{k1}
\end{align*}
and, hence,
\begin{align}
c^{(1)}_k&= \frac{V^{(0)}_{k1}}{C_k(1-\lambda_k)}, \quad k=2,\dots,K. \label{ccoeff}
\end{align}
This demonstrates that the steady-state solution is particularly sensitive to perturbations if many eigenvalues of the unperturbed flow are close to unity, since
\begin{equation}
\vec{u}^{(1)}_1=\sum_{k=2}^K \frac{V^{(0)}_{k1}}{C_k(1-\lambda_k)} \vec{u}_k\label{u1'}.
\end{equation}

In full analogy to the first-order correction terms $u^{(1)}_1$ and $\lambda^{(1)}_1$, we can determine the second-order correction terms $u^{(2)}_1$ and $\lambda^{(2)}_1$. Specifically, defining 
\begin{equation}
V^{(1)}_{kl}:=\vec{u}^*_k \mathbf{V} \vec{u}^{(1)}_l
\end{equation}
\noindent
and multiplying Eq.~$\eqref{zweitekorrektur}$ from the left by $\vec{u}^*_1$, we obtain
\begin{align*}
\vec{u}^*_1 \mathbf{A} \vec{u}^{(2)}_1 + \vec{u}^*_1 \mathbf{V} \vec{u}^{(1)}_1 &= \vec{u}^*_1 \cdot \vec{u}^{(2)}_1 + \vec{u}^*_1 \lambda^{(1)}_1 \vec{u}^{(1)}_1 + \vec{u}^*_1 \lambda^{(2)}_1 \vec{u}_1,
\end{align*}
which yields
\begin{align}
V^{(1)}_{11} &= \lambda^{(1)}_1 \sum_{k=2}^K c^{(1)}_k \vec{u}^*_1 \cdot \vec{u}_k + \lambda^{(2)}_1 \vec{u}^*_1 \cdot \vec{u}_1 = C_1 \lambda^{(2)}_1.
\end{align}
Similar as for the first-order correction, we thus find:
\begin{align}
\lambda^{(2)}_1 &= -\frac{f_k}{C_1} u_{1,k}^{(1)}.
\end{align}
\noindent
For the associated second-order correction to the eigenvector, we again make the ansatz
\begin{align}
u^{(2)}_1 &= \sum_{k=2}^K c^{(2)}_k \vec{u}_k
\end{align}
\noindent
and multiply $\vec{u}^*_k$ from the left to determine the unknown expansion coefficients:
\begin{align*}
\vec{u}^*_k \mathbf{A} \vec{u}^{(2)}_1 &+ \vec{u}^*_k  \mathbf{V} \vec{u}^{(1)}_1 = \\ & \vec{u}^*_k \cdot \vec{u}^{(2)}_1 + \lambda^{(1)}_1 \vec{u}^*_k \cdot \vec{u}^{(1)}_1 + \lambda^{(2)}_1 \vec{u}^*_k \cdot \vec{u}_1.
\end{align*}
\noindent
Inserting the expansions for $u^{(2)}_1$ and $u^{(1)}_1$ yields
\begin{align*}
&\sum_{l=2}^K c^{(2)}_l \vec{u}^*_k \mathbf{A} \vec{u}_l + V^{(1)}_{k1} = \\
&\qquad \sum_{l=2}^K c^{(2)}_l \vec{u}^*_k \cdot \vec{u}_l + \lambda^{(1)}_1 \sum_{l=2}^K c^{(1)}_l \vec{u}^*_k \cdot \vec{u}_l + \lambda^{(2)}_1 C_1 \delta_{k1},
\end{align*}
\noindent 
and we obtain
\begin{align}
c^{(2)}_k &= \frac{V^{(1)}_{k1}-\lambda^{(1)}_1 c^{(1)}_k}{C_k(1-\lambda_k)}.
\end{align}
\noindent
As for the first-order correction, this implies that also the impact of the second-order correction is the larger, the closer the eigenvalues $\lambda_k$ are to unity.

Taken together, the perturbed first eigenvalue and the associated eigenvector read
\begin{align}
\tilde{\vec{u}}_1 = \vec{u}_1 &+ \sigma \sum_{k=2}^K \frac{V^{(0)}_{k1}}{C_k(1-\lambda_k)} \vec{u}_k \nonumber \\
&+ \sigma^2 \sum_{k=2}^K \frac{V^{(1)}_{k1}-C_k\lambda^{(1)}_1 c^{(1)}_k}{C_k(1-\lambda_k)} \vec{u}_k+\mathcal O(\sigma^3)\\
\tilde{\lambda}_1 &= \lambda_1 - \sigma \frac{f_k}{C_1} u_{1,j} - \sigma^2 \frac{f_k}{C_1} u^{(1)}_{1,j} + \mathcal O(\sigma^3).
\end{align}
In a similar way, we can express further higher-order correction terms $\lambda^{(\alpha)}_1$ and $\vec{u}^{(\alpha)}_1$ with expansion coefficients $c^{(\alpha)}_k$ yielding
\begin{align}
\lambda^{(\alpha)}_1 &= \frac{V^{(\alpha-1)}_{11}}{C_1}:=\frac{1}{C_1} \vec{u}^*_1 \mathbf{V} \vec{u}_1^{(\alpha-1)},\\
\vec{u}^{(\alpha)}_1 &= \sum_{k=2}^K c^{(\alpha)}_k \vec{u}_k
\end{align}
\noindent
with
\begin{align}
c^{(\alpha)}_k &= \frac{1}{C_k(1-\lambda_k)}\left(V^{(\alpha-1)}_{k1} - C_k \sum_{l=1}^{\alpha-1} \lambda^{(l)}_1 c_k^{(\alpha-l)}\right).
\end{align}

The aforementioned considerations apply to the case of a fully localized absorption of particles at a single node. The more general case of spatially distributed absorption can be treated in an analogous way by specifying the matrix elements of $\mathbf{V}$ accordingly.

\subsection{Absorption problem with constant supply}
\label{sec:const_input}

In the previous section, we have considered a case where the perturbed system has been efficiently leaked by successive absorption of particles, so that its asymptotic state is characterized by an absence of particles. The considered perturbation-theoretic approach described how the steady state of the unperturbed system is modified, depending on the eigenvalues of the unperturbed flow. In the following, we will focus on the steady state of the perturbed system in a setting where the absorption is compensated by an additional supply of particles.

Here, we consider the case of constant non-zero particle supply, i.e., $s_j\ge 0$ $\forall j\in\{1,\dots,K\}$ and there exists at least one $j$ with $s_j>0$. Being interested in the steady state of the perturbed system, we first recall that Eq.~\eqref{gestoertessystem} has a steady-state solution if and only if $\left(\mathbf{1}-\tilde{\mathbf{A}}\right)$ is invertible. This invertibility holds if and only if $\tilde{\mathbf{A}}$ has no eigenvalue of one. Otherwise, for sufficiently long time $t$, the number of particles would increase linearly with time proportional to the associated right eigenvector
\begin{equation*}
\vec{n}(t)\propto\tilde{\vec{u}}_1 t.
\end{equation*}
\noindent
In the presence of an additional matrix $V$ describing the absorption of particles, the largest eigenvalue of $\tilde{\mathbf{A}}$ can be smaller than unity, which guarantees the invertibility of $\left(\mathbf{1}-\tilde{\mathbf{A}}\right)$ and, hence, the existence of a steady state as given in Eq.~\eqref{steadystate}.

In order to express the eigenvalues and eigenvectors of $\tilde{\mathbf{A}}$, we again use a perturbation-theoretic approach by expanding all eigenvalues and eigenvectors into powers of $\sigma$. Assuming non-degeneracy, we obtain
\begin{align*}
\lambda^{(1)}_i &= -\frac{f_k}{C_i}u_{i,k} \sum_{j=1}^K u^*_{i,j} A_{j} \\
\vec{u}^{(1)}_i &= \sum_{j\neq i} \left(\frac{ V^{(0)}_{ji}}{C_j(\lambda_i-\lambda_j)}\right) \vec{u}_j
\intertext{for the first-order correction term and}
\lambda^{(2)}_i &= -\frac{f_k}{C_i} \left(u_{i,k}^{(1)}-\frac{1}{C_i}u_{i,k} \right)\sum_{j,l=1}^K u^*_{i,j} A_{jk} \\
\vec{u}^{(2)}_i &= \sum_{j\neq i} \left(\frac{ V^{(1)}_{ji} - V_{ii}^{(0)} \frac{C_j}{C_i} c_j^{(1)}}{C_j(\lambda_i-\lambda_j)}\right) \vec{u}_j
\end{align*}
for the second-order correction term.

Let us suppose that we know the eigenvalues and eigenvectors of $\tilde{\mathbf{A}}$, which again form a basis of $\mathbb{R}^K$. Thus, we can write
\begin{equation}
\vec{n}_{st} = \sum_{i=1}^K a_i\tilde{\vec{u}}_i \quad\text{and}\quad \vec{s}=\sum_{i=1}^K b_i \tilde{\vec{u}}_i.
\end{equation}
\noindent
Inserting this into Eq.~$\eqref{steadystate}$ yields
\begin{align}
\sum_{i=1}^K a_i \tilde{\vec{u}}_i &= \left(\mathbf{1}-\tilde{\mathbf{A}}\right)^{-1}\sum_{i=1}^K b_i \tilde{\vec{u}}_i,\nonumber
\end{align}
which is equivalent to
\begin{align}
\sum_{i=1}^K a_i \tilde{\vec{u}}_i(1-\tilde\lambda_i) &= \sum_{i=1}^K b_i \tilde{\vec{u}}_i.\nonumber
\end{align}
This is true, if and only if
\begin{align}
a_i&=\frac{b_i}{1-\tilde\lambda_i}, \quad i=1,\dots,K.\label{coeffizientsteady}
\end{align}
Thus, for a given source term $s$ we obtain the steady state
\begin{align}
\vec{n}_{st} &= \sum_{i=1}^K \frac{b_i}{1-\tilde\lambda_i} \tilde{\vec{u}}_i \quad \mbox{with} \quad b_i=\tilde{\vec{u}}_i^*\cdot\vec{s}.
\end{align}


\subsection{Modified transition matrix with probability conservation}
\label{sec:conserve}

Finally, we consider the case where the transition matrix $\mathbf{A}$ is perturbed such as the probability is conserved, which implies $\vec{s}=\vec{0}$ and 
$$\sum_{i=1}^K \tilde{A}_{ij} = \sum_{i=1}^K (A_{ij}+\sigma V_{ij})=1$$
for all $j$, i.e., $\sum_m V_{mk} = 0$ as shown in Ref.~\cite{f13b}.

Here, we analytically express the steady state $\tilde{\vec{u}}_1$ of the perturbed matrix $\tilde{\mathbf{A}}$. Since $\tilde{\mathbf{A}}$ is now itself a column-stochastic matrix, $\tilde{\vec{u}}_1$ is the right eigenvector belonging to the eigenvalue $\tilde{\lambda}_1$ and describes the steady-state distribution. As in the previous cases, we express the first-order correction term as a linear combination of the eigensystem of the unperturbed case as 
\begin{align}
\tilde{\vec{u}}_1 = \vec{u}_1 + \sigma \sum_{i\neq 1} d_i \vec{u}_i
\label{eq:exp_consv}
\end{align}
and make use of
\begin{equation}
\tilde{\lambda}_1 \tilde{\vec{u}}_1 = \tilde{\vec{u}}_1 = (\mathbf{A}+\sigma\mathbf{V}) \tilde{\vec{u}}_1.
\label{eq:eig_consv}
\end{equation}
Substituting Eq.~(\ref{eq:exp_consv}) into Eq.~(\ref{eq:eig_consv}) and
multiplying with $\vec{u}^*_k$ from the left, we obtain
\begin{align*}
\delta_{k1} + \sigma \sum_{i=2}^K d_i \delta_{ki} &= \lambda_k \left(\delta_{k1}+\sigma \sum_{i=2}^K d_i \delta_{ki}\right) + \\
&\quad \quad \sigma \left(V^{(0)}_{k1} + \sigma\sum_{i=2}^K d_i V^{(0)}_{ki}\right),\nonumber
\end{align*}
which implies
\begin{align}
d_k &= \frac{V^{(0)}_{k1} + \sigma\sum_{i=2}^K d_i V^{(0)}_{ki}}{C_k(1-\lambda_k)}, \quad k=2,\dots,K.\label{coefficients}
\end{align}
Note that the latter expression is exact, but hard to evaluate in practice. Letting $\sigma\to 0$, we obtain a first-order correction
\begin{equation}
d_k^{(1)} = \frac{V_{k1}^{(0)}}{C_j(1-\lambda_k)},
\end{equation}
\noindent
which is equivalent to the result of Eq.~\eqref{ccoeff} for the absorption problem.

\section{Numerical example}
\label{sec:examples}

\begin{figure}[htbp]
\vspace*{2mm}
\begin{center}
\includegraphics[width=8.2cm]{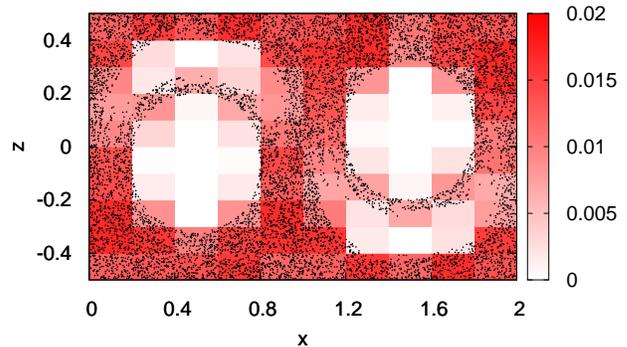}
\end{center}
\caption{\label{fig1} 
(Color online) Steady-state distribution of particles in the Lagrangian chaos model described by Eq.~(\ref{eq:lagrange}).
Black dots indicate the position of 10,000 particles with randomly distributed initial conditions after $t=20$. Numerical integration was performed with a fourth-order Runge-Kutta method with $\Delta t = 0.001$. The background color represents the residence probability of particles at each node.}
\end{figure}

\begin{figure*}[th]
\vspace*{2mm}
\begin{center}
\begin{minipage}{0.45\hsize}
\includegraphics[width=\textwidth]{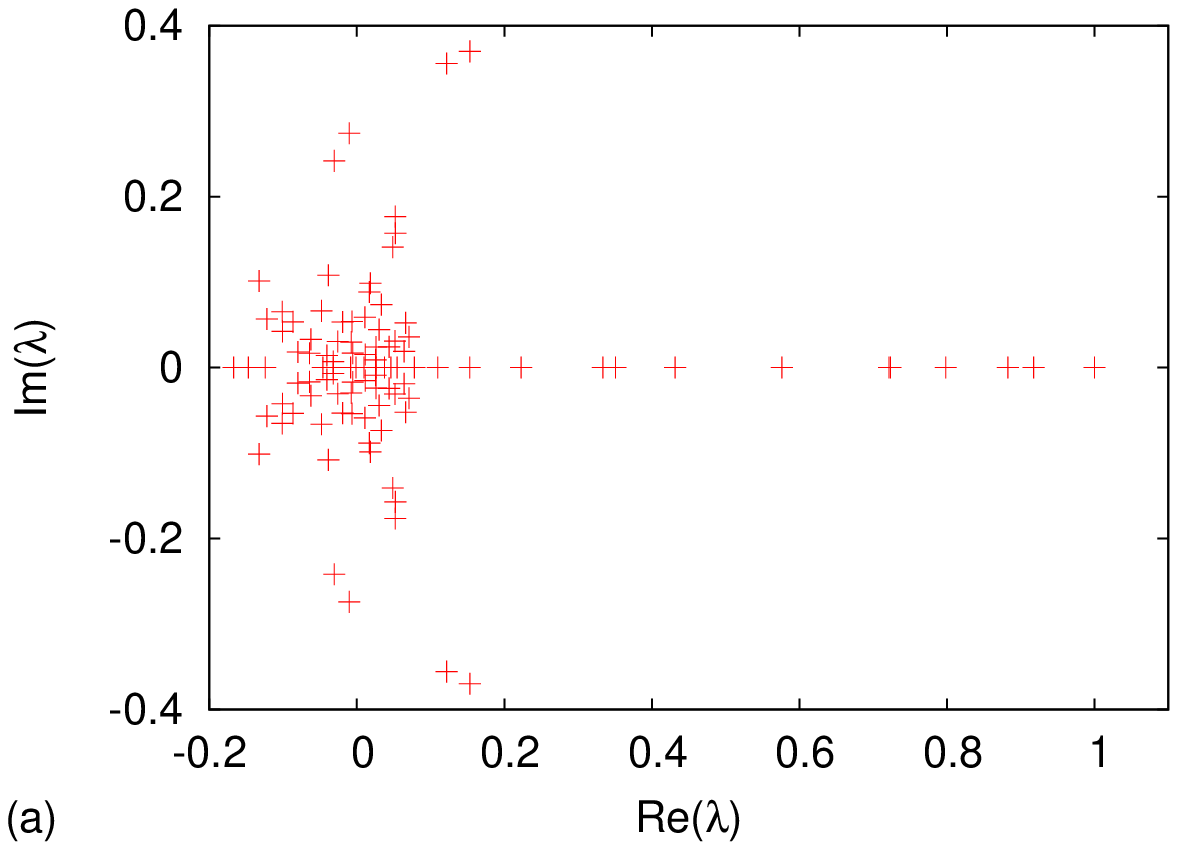}
\end{minipage}
\begin{minipage}{0.45\hsize}
\includegraphics[width=\textwidth]{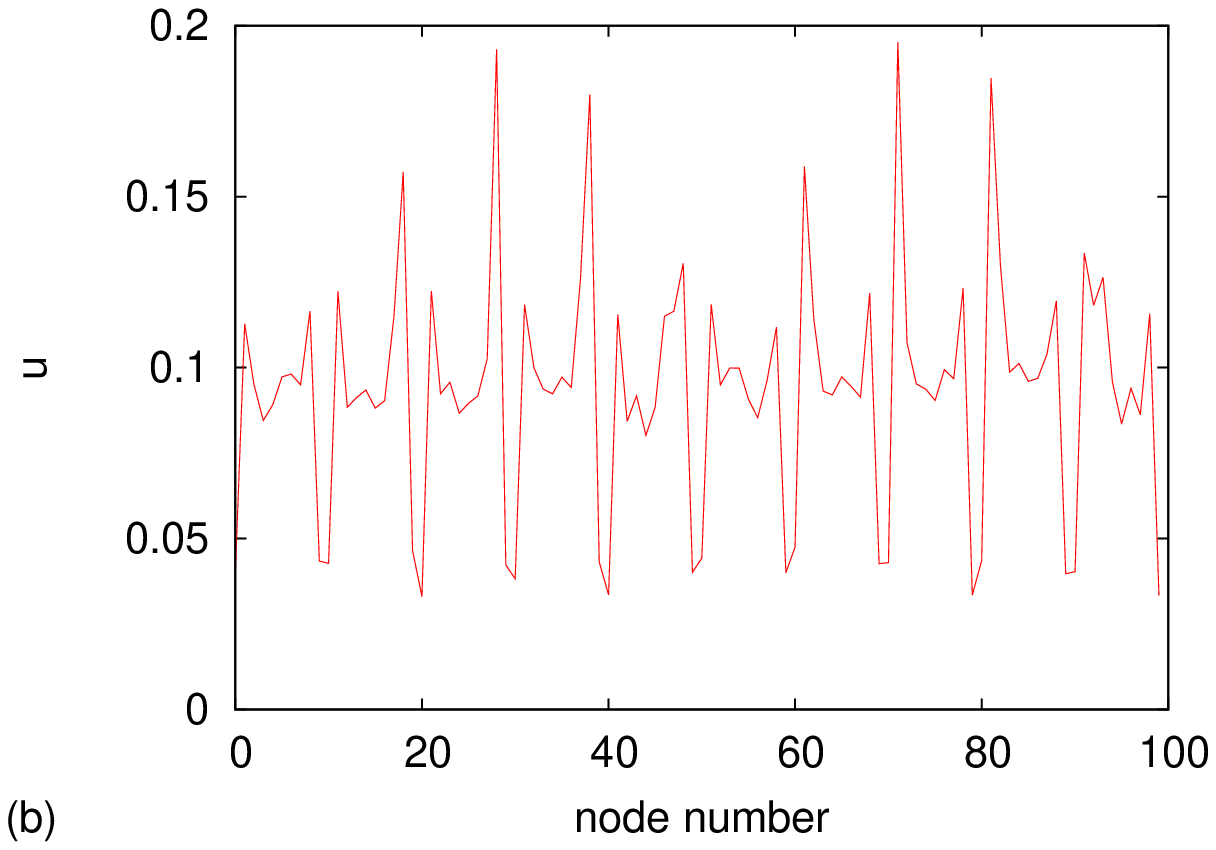}
\end{minipage}\\
\begin{minipage}{0.45\hsize}
\includegraphics[width=\textwidth]{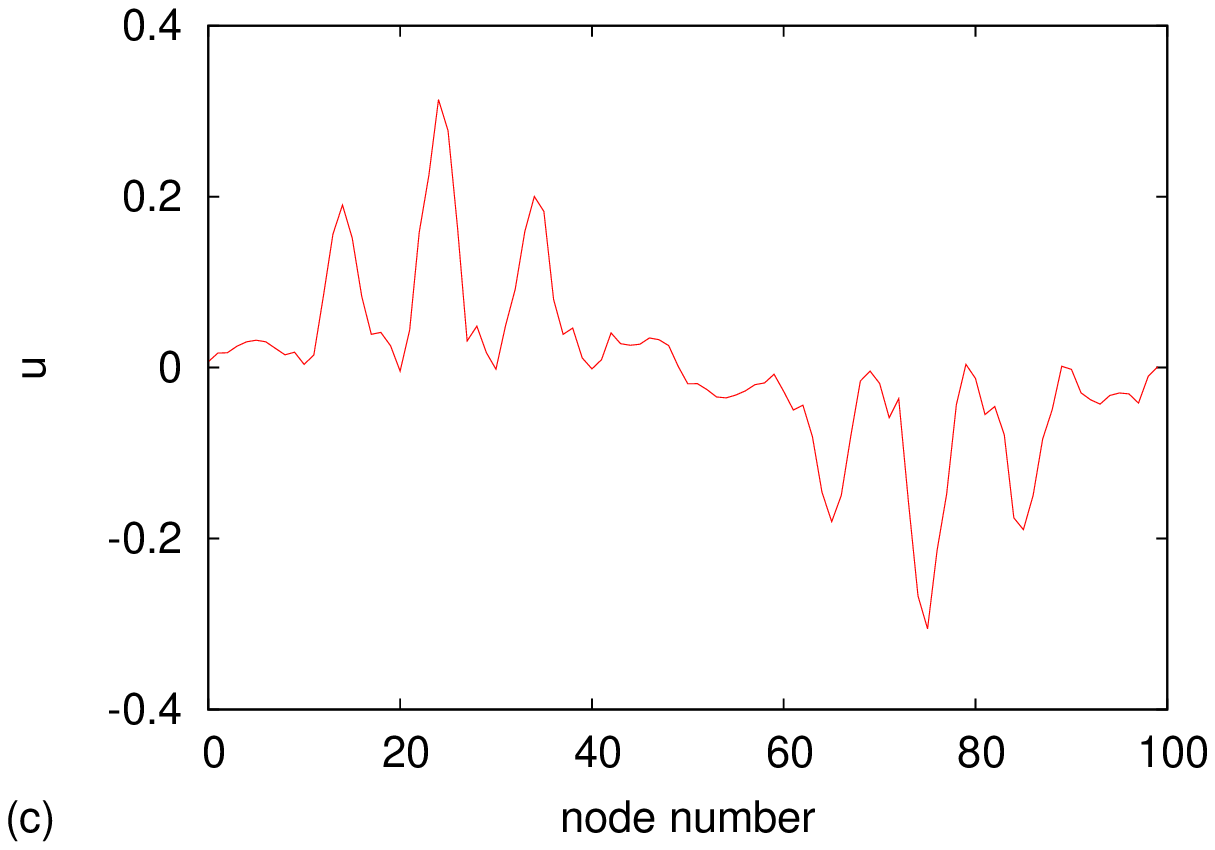}
\end{minipage}
\begin{minipage}{0.45\hsize}
\includegraphics[width=\textwidth]{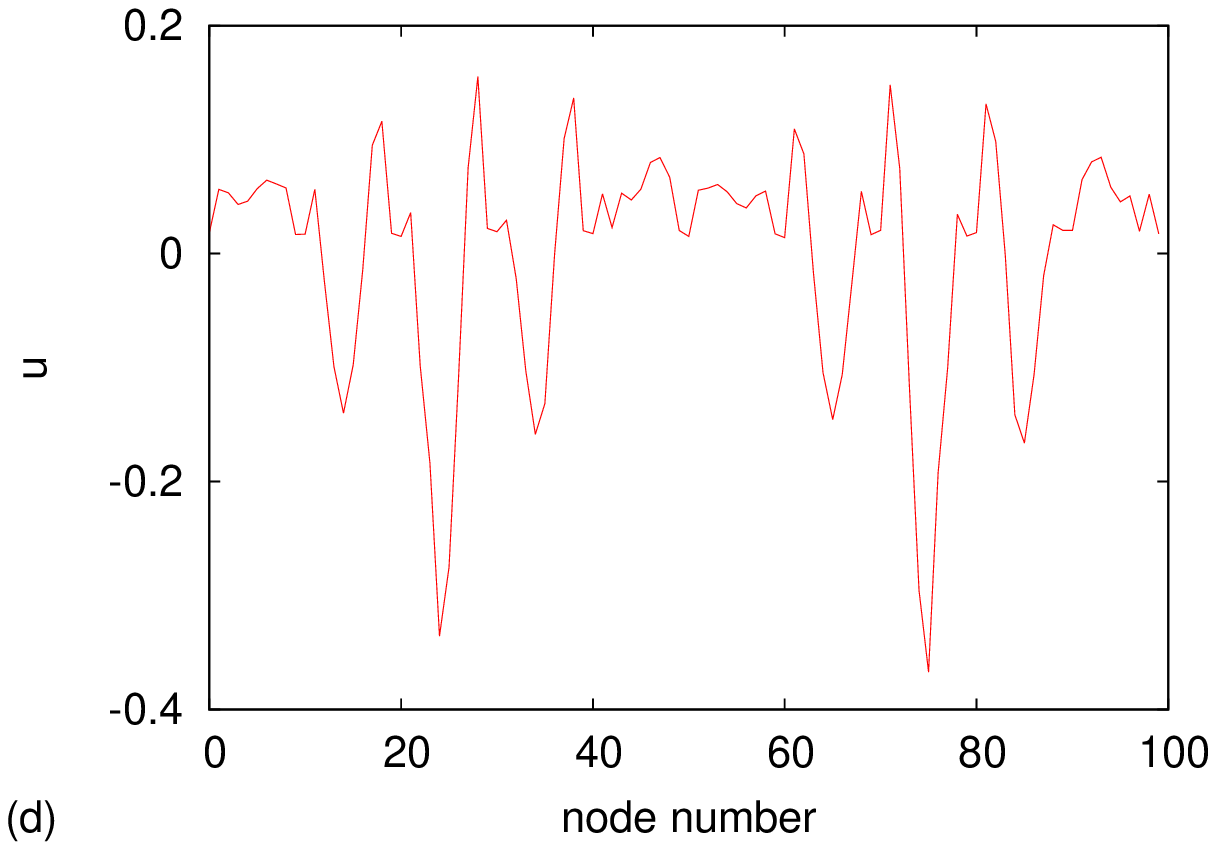}
\end{minipage}
\end{center}
\caption{\label{fig4} 
(Color online) (a) Eigenvalues of the transition matrix on the complex plane for the Lagrangian chaos model. There exists one unique eigenvalue at unity, corresponding to the conservation of probability. The absolute values of all other eigenvalues are smaller than 1. (b-d) Components of the first (b), second (c), and third (d) right eigenvectors of the system. Note that since the corresponding eigenvalues are real, these eigenvectors have exclusively real entries.
}
\end{figure*}

To illustrate the framework proposed in the previous section, we numerically study a time-dependent two-dimensional velocity field model exhibiting Lagrangian chaos~\cite{omhm91}. Here, the passive advection of particles starting at time $t$ at a position $(x,z)$ is described by a stream function as 
\begin{align}
\frac{dx}{dt}(x,z,t) = - \frac{\partial \Psi}{\partial z},\ 
\frac{dz}{dt}(x,z,t) = \frac{\partial \Psi}{\partial x},
\label{eq:lagrange}
\end{align}
where
\begin{align}
\Psi(x,z,t) &=
\frac{a}{\pi} \sin[\pi\{ x+b\sin(2\pi t)\} ] W(z)
\end{align}
with $a=3.1$, $b=0.0404411$ and
\begin{multline}
W(z) = \cos(q_0 z) - \alpha \cosh(q_1 z) \cos(q_2 z) \\
+ \beta \sinh(q_1 z) \sin (q_2 z).
\end{multline}
The considered system is periodic in both time and space as $\Psi (x+2,z,t) = \Psi (x,z,t)$ and $\Psi (x,z,t+1)=\Psi(x,z,t)$. The latter periodicity enables us to set the time step of the transition matrix as $\tau=1$ to effectively treat the problem like for a stationary flow.
For convenience, we constrain the time derivative of $z$ as $dz/dt < 0$ at $ z \lesssim 1/2$ and $dz/dt > 0$ at $z \gtrsim -1/2$, so that particles are forced to stay within the domain $-1/2<z<1/2$. The resulting flow pattern is shown in Fig.~\ref{fig1}, exhibiting two roll-like structures resembling a two-dimensional projection of Rayleigh-B\'enard convection. We divide the whole system into 100 sub-regions of size $0.2\times 0.1$, which are interpreted as the nodes of a Lagrangian flow network. The steady state solution of the system is approximated by
the right eigenvector of the transition matrix corresponding to the eigenvalue 1 shown in Fig.~\ref{fig4}b.

Let us now consider the absorption problem discussed in Sec.~\ref{sec:absorption}.
We remove a particle when it reaches a certain region of the flow described by an absorbing node $k$ with the probability $g_k$. Here, we consider the extreme case of $g_k=1$, i.e., all particles are absorbed at the considered node. This absorption takes place every $\tau=1$ time steps. After an initial transient, we expect that the number of particles decays exponentially (Eq.~\ref{eq:decay_rate}). 

\begin{figure}[htbp]
\vspace*{2mm}
\begin{center}
\includegraphics[width=4.1cm]{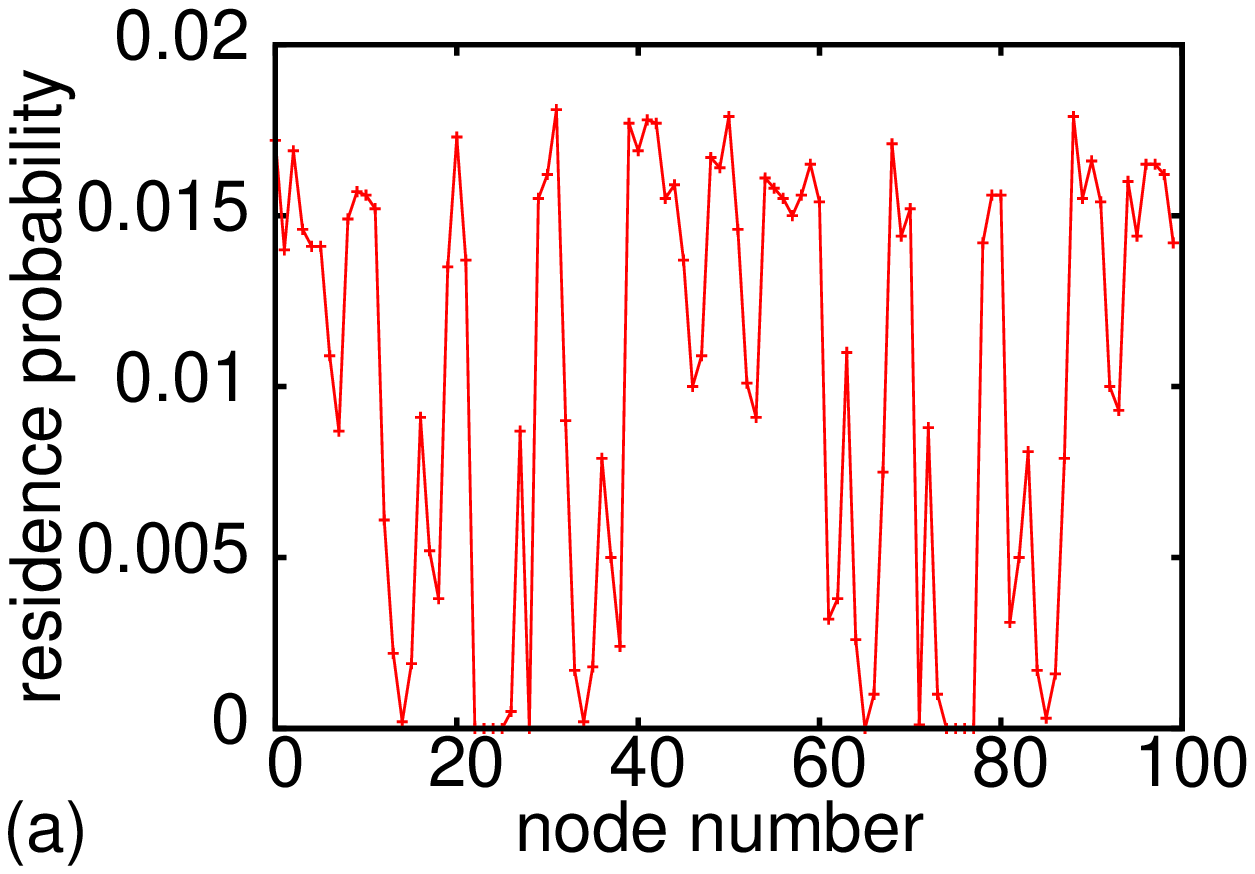}
\includegraphics[width=4.1cm]{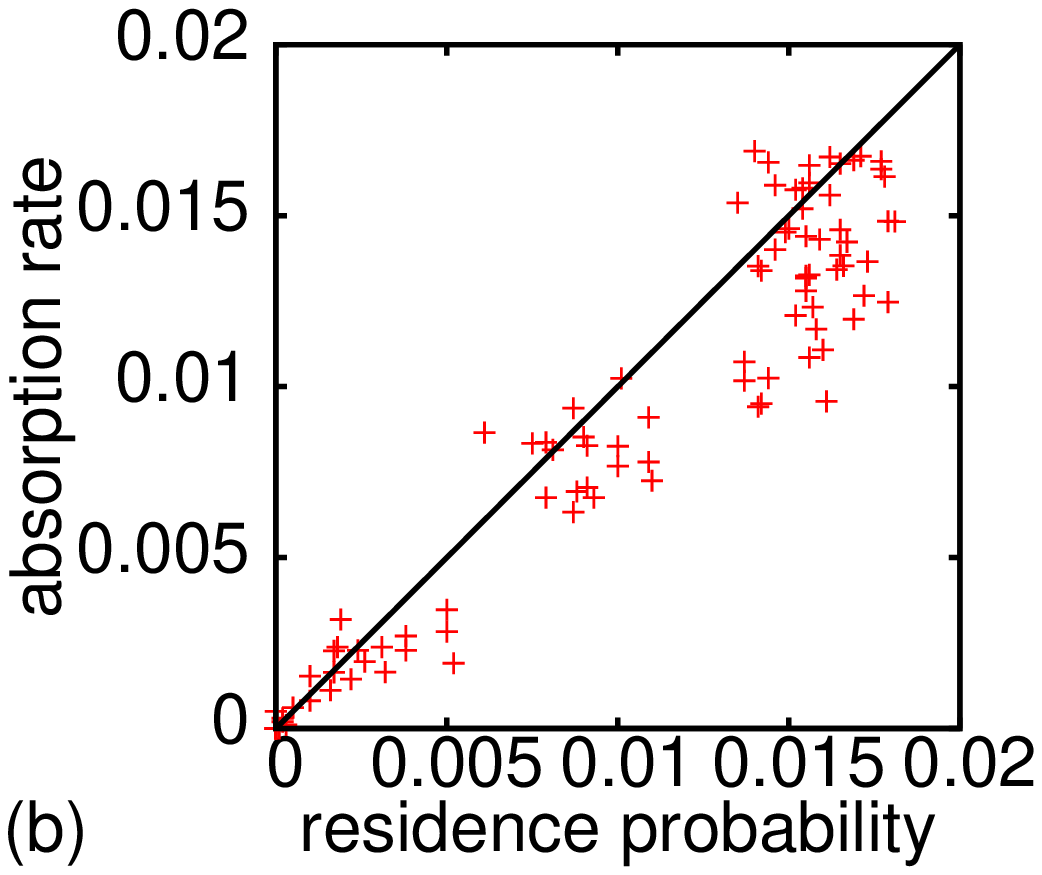}
\end{center}
\caption{\label{fig2} 
(Color online) (a) Residence probability of particles at each node in the absence of absorption. (b) Absorption rate $\tilde \lambda_1$ and residence probability at the absorption node for the absorption problem in the Lagrangian chaos model.  
}
\end{figure}

Let $\tilde \lambda_1$ again be the largest eigenvalue of the perturbed transition matrix (Eq.~\ref{eq:pert_transition}), which coincides with the absorption rate of the considered problem. As predicted by our previous analytical considerations, this rate depends on the absorption node $k$. Equation (\ref{eq:1steig_fk}) suggests that $\tilde \lambda_1$ is approximated by the residence probability of the absorption node (Fig.~\ref{fig2}(a)) at the first order of the perturbation expansion. Indeed, Fig.~\ref{fig2}(b) indicates that both characteristics exhibit a rather strong correlation. Deviations from a one-by-one correspondence can be attributed to the coarse-graining of physical space underlying the network approximation, still too small particle numbers and the ignorance with respect to higher-order terms in the perturbation expansion. In particular, we emphasize that the co-existence between regular (periodic and quasi-periodic) and chaotic domains is an inherent characteristic of the considered two-dimensional Lagrangian chaos model as well as a great variety of similar systems. The clear separation between these domains is partially relieved by our coarse-graining, implying that while particles cannot migrate between domains of different dynamics, the transition matrix may exhibit non-zero transition probabilities since individual fluid volumes may cover parts of different domains. As a consequence, the leading eigenvector (Fig.~\ref{fig4}b) of the coarse-grained transition matrix may not provide a sufficient approximation of the residence probabilities of the system (Fig.~\ref{fig2}a). Future studies should therefore address the conditions for the validity of the corresponding approximations in further detail.

%
%
%
%

%
%

%
%

\section{Conclusions}  
\label{sec:conclusion}

We have considered the problem of coarse-graining passive advection processes in velocity fields by means of a Markov chain approach utilizing the discretized Perron-Frobenius operator, which can be interpreted as a Lagrangian flow network. Specifically, we have studied how the eigensystem of the corresponding transition matrix can be approximated by perturbation-theoretic methods in the case of three different types of perturbations: (i)~localized absorption, (ii)~absorption in the presence of constant particle supply, and (iii)~changes of the velocity field (i.e., keeping the total probability mass conserved). We have shown that in all three situations, the correction terms to the original eigenvalues and eigenvectors of the transition matrix can be analytically expressed up to arbitrarily high order. Specifically, our analysis reveals that modifications due the system's steady state imposed by any of the aforementioned types of perturbations can be expressed in terms of the eigensystem of the unperturbed flow.

Our method is motivated by common perturbation expansions in other areas of science like quantum mechanics, but mostly focuses on the properties of the system's steady-state solution which can exist in case of a probability-conserving perturbation (e.g., a change of the underlying flow or local absorption compensated by a constant particle supply). However, the techniques used in our study can be equally employed for studying modifications to higher-order variability modes as expressed by the transition matrix' eigentriples of higher order. Since the spectral representation of the Perron-Frobenius operator is a widely used tool in the mathematical analysis of fluid dynamical problems, we are confident that the results are of potential use in a variety of corresponding applications.

To this end, most of our presented considerations have been purely theoretical. It will be a subject of future studies to provide further numerical verification of the obtained results for different types of flow patterns and perturbations. Specifically, we have demonstrated that the correction terms to the unperturbed situation are the more important, the closer the eigenvalues of the unperturbed system are to unity. This result allows generally characterizing the vulnerability of transportation processes depending on the spectral properties of the underlying flow pattern and classifying real-world transportation problems accordingly.

Since the method presented in this paper is capable of predicting the response of the steady state of the passive advection problem as a generic type of transport phenomenon, there are many possible applications of the developed expressions in various scientific disciplines. One example relying on some previous work by one of the authors~\cite{fkd11,fkd16} are communication patterns related to the flow of mobile agents. Over the last years, it has become possible to track the motion of individuals using GPS or call detail records from mobile devices such as cell phones, which allows constructing a transition matrix describing the commuter dynamics between different urban sub-regions and thus gaining information on human mobility and activity patterns~\cite{gonzalez2008understanding}. Practical applications of such an urban commuter transition matrix include the study of infectious disease spreading, urban planning, design of evacuation plans in case of disasters, etc. Notably, for such applications, the perturbation-theoretic approach would call for an application to more than just the leading eigentriple of the transition matrix.

Finally, we emphasize the relevance of the developed approach for studying contamination processes in general, like the spreading of oil spills or volcanic aerosols, or certain geoengineering proposals related to injection of reflecting or absorbing particles into different layers of the Earth's atmosphere to counteract global warming. In all these real-world cases, the time-dependence of the underlying flow will become particularly relevant. In our analytical considerations presented in this work, we have exclusively considered stationary flows; however, the perturbation-theoretic approach should be easily extendable to cases operating with time-dependent transition matrices and, hence, time-dependent eigentriples. Additional questions that have been left open intentionally in this initial study include the explicit treatment of degenerate eigenvalues of the transition matrix as well as the perturbation-theoretic expansion of topological characteristics of the associated flow networks like (in/out-) degree and strength, local clustering coefficient and others. We leave a detailed investigation of these problems as subjects of future work.

\subsection*{Acknowledgements}
J.F.D. acknowledges financial support by the BMBF project GLUES, the Stordalen Foundation (via the Planetary Boundary Research Network PB.net) and the Earth League's EarthDoc program. 
R.V.D. received funding by the German Federal Ministry for Education and Research (BMBF) via the BMBF Young Investigator's Group CoSy-CC$^2$ (``Complex Systems Approaches to Understanding Causes and Consequences of Past, Present and Future Climate Change, grant no.~01LN1306A'') and the IRTG 1740 ``Dynamical Phenomena in Complex Networks'' jointly funded by DFG and FAPESP. 
The authors acknowledge valuable input from discussions with Kazuyuki Aihara and J\"urgen Kurths.

\bibliographystyle{unsrt}
\bibliography{naoya_bib}

\end{document}